\documentstyle[12pt]{article}

\begin{document}
\begin{flushright}
\baselineskip=12pt
CTP-TAMU-07/97\\
DOE/ER/40717--39\\
ACT-03/97\\
\tt hep-ph/9702237
\end{flushright}

\begin{center}
\vglue 1.5cm
{\Large\bf M-theory Inspired No-scale Supergravity}
\vglue 2.0cm
{\Large Tianjun Li$^{1,2}$, Jorge L. Lopez$^3$, and D.V. Nanopoulos$^{1,2}$}
\vglue 1cm
\begin{flushleft}
$^1$Center for Theoretical Physics, Department of Physics, Texas A\&M
University\\ College Station, TX 77843--4242, USA\\
$^2$Astroparticle Physics Group, Houston Advanced Research Center (HARC)\\
The Mitchell Campus, The Woodlands, TX 77381, USA\\
$^3$ Bonner Nuclear Lab, Department of Physics, Rice University\\ 6100 Main
Street, Houston, TX 77005, USA\\
\end{flushleft}
\end{center}

\vglue 1.5cm
\begin{abstract}
We propose a supergravity model that contains elements recently shown to arise in the strongly-coupled limit of the $E_8\times E_8$ heterotic string (M-theory), including a no-scale--like K\"ahler potential, the identification of the string scale with the gauge coupling unification scale, and the onset of supersymmetry breaking at an intermediate scale determined by the size of the eleventh dimension of M-theory. We also study the phenomenological consequences of such scenario, which include a rather constrained sparticle spectrum within
the reach of present-generation particle accelerators. 
\end{abstract}

\vspace{0.5cm}
\begin{flushleft}
\baselineskip=12pt
January 1997\\
\end{flushleft}
\newpage
\setcounter{page}{1}
\pagestyle{plain}
\baselineskip=14pt

Elementary particle phenomenology based on conventional weakly-coupled string theory has met with considerable success over the
years. These successes include the taming of quantum gravity, the prediction
of space-time supersymmetry (at least at the Planck scale), the prediction of realistic gauge groups and matter representations, and the illumination of the family and fermion mass problems. Despite these accomplishments, string theory
has not been able to deliver on its original promise of an unambiguous description of high-energy physics in terms of a single parameter: the Planck mass. This problem is traced to the very large degeneracy of the string vacuum. In addition, different string theories ({\em e.g.}, heterotic, Type I, Type II)
are also allowed. These and other difficulties, like the supersymmetry
breaking mechanism, were thought to require an understanding of the strongly-coupled limit of string. Of course, if strings are strongly coupled
much doubt is cast on the results obtained at weak coupling that are not
protected by non-renormalization theorems. One such result is the desirable prediction of a no-scale supergravity structure \cite{no-scale} with zero vacuum energy at tree level, at least in some string compactifications
\cite{Witten85,LNY+LN}.

The recent discovery of extended duality symmetries relating different string theories at various values of the string coupling \cite{Polchinski} has provided a much needed tool to study the strongly-coupled limit of strings. These new developments have ellucidated some of the problematic features of weakly-coupled strings \cite{Ferrara}, such as the relation between different string theories, a new mechanism to reconcile the gauge coupling and string unification scales, and a possible topological mechanism for supersymmetry breaking. The new symmetries require an enlarged set of possible theories, including five-branes and 11-dimensional supergravity.

In this paper we concentrate on the low-energy supersymmetry phenomenology
that might arise in the strongly-coupled limit of the $E_8\times E_8'$ heterotic string in ten dimensions. This theory has been shown \cite{HW} to map into a weakly-coupled eleven-dimensional ``M-theory", which has
as its low-energy limit eleven-dimensional supergravity. The eleventh dimension ($\rho$) has an orbifold structure ($S^1/Z_2$): at one end
live the observable fields contained in $E_8$, at the other end live the hidden sector fields contained in $E_8'$, and in the middle (`bulk') propagate the
gravitational fields. One can distinguish a few mass scales in such scenario:
the eleven-dimensional Planck scale ($M_{11}$), the size of the eleventh
dimension ($\rho$), and the compactification mass scale ($R^{-1}$). Strong-weak
coupling duality relates these three scales as follows \cite{Witten,BD,AQ}:
\begin{eqnarray}
M_{11}&=&(2\alpha_G\, V)^{-1/6}\ ,
\label{eq:M11def}\\
\rho^{-1}&=&\left({2\over\alpha_G}\right)^{3/2}M^{-2}_{\rm Pl}\, V^{-1/2}\ ,
\label{eq:rhodef}\\
V&=& R^d\, r^{6-d}\,;\quad 0<d\le6\,,
\label{eq:Vdef}
\end{eqnarray}
where $\alpha_G$ is the unified gauge coupling ($\alpha_G\approx{1\over25}$
as deduced from running the Standard Model gauge couplings), 
$M_{\rm Pl}\approx10^{19}\,{\rm GeV}$ is the usual Planck mass, and $V$ is the
volume of the compactified space (which does not include the eleventh dimension). This volume depends on the compactification manifold used, and
in Eq.~(\ref{eq:Vdef}) it has been parametrized in terms of two internal
radii: $R$ representing $d$ possibly `large' dimensions and $r$ representing
$6-d$ possible `small' dimensions. In compactification on symmetric Calabi-Yau
manifolds $d=6$. 

The relevance of these equations to gauge coupling unification was pointed out by Witten \cite{Witten}, who noticed that $M_{11}$, essentially the ``string"
unification scale ($M_{\rm string}$), might be lowered down to the gauge coupling unification scale $M_{\rm LEP}$ by choosing a sufficiently large compactification volume $V$. Previously the only way to reduce $M_{\rm string}$
(at tree level) was to increase the unified gauge coupling, which then disagreed with the result obtained by running the gauge couplings up from low
energies.\footnote{Earlier attempts at reconciling $M_{\rm string}$ with $M_{\rm LEP}$ are reviewed in Ref.~\cite{Dienes}.} 

A very interesting result that also emerges from `M-phenomenology' is the
appearance of the no-scale supergravity structure \cite{BD,Horava}.\footnote{An
explicit calculation, along the lines of Ref.~\cite{Witten85}, also supports this conclusion \cite{Tianjun}.} No-scale supergravity is then seen to emerge both in the weakly- and in the strongly-coupled limits of the heterotic string. In particular, the K\"ahler potential seems to be of the form that guarantees universality of the scalar masses ({\em i.e.}, $m_0=0$) at the scale where supersymmetry-breaking effects are first felt in the observable sector ($\Lambda_{\rm susy}$). A more model-dependent question (as it involves the superpotential) is the form of the parameter $A_0$, which for simplicity we will take to vanish ($A_0=0$), as is typical in weakly-coupled string realizations of no-scale supergravity \cite{no-scale,Witten85}. Even more model dependent are the values of the Higgs mixing parameter $\mu$ and its associated supersymmetry breaking parameter $B_0$. These will be left unspecified in our present analysis. The gaugino masses will be taken to be universal ($m_{1/2}$) and will parametrize the low-energy spectrum. These may be obtained in principle from knowledge of the gauge kinetic function and the gravitino mass.
After using the radiative electroweak symmetry breaking conditions
at the electroweak scale, we need to specify only two parameters to fully
describe the low-energy supersymmetry spectrum: $m_{1/2}$ and the ratio of Higgs vacuum expectaction values $\tan\beta$.

Also of great interest is the mechanism of supersymmetry breaking that appears
to be emerging in M-phenomenology. Horava \cite{Horava} has argued that supersymmetry breaking, say via gaugino condensation in the hidden sector, is not felt immediately in the observable sector because of a topological obstruction (essentially the eleventh dimension of length $\rho$ that separates the two sectors). Supersymmetry breaking becomes apparent only after the renormalization scale is low enough to not reveal the presence of the eleventh
dimension anymore ({\em i.e.}, when this extra dimension becomes `compact').
In practice one is to allow for non-vanishing supersymmetry breaking
parameters only for scales $Q<\rho^{-1}$. This effect can leave a deep imprint
on the low-energy sparticle spectrum, which depends quantitatively on the
amount of `running' of these parameters. As has been noted in connection with
the relevance of gaugino condensation \cite{Horava,AQ} (and as we emphasize below), $\rho^{-1}$ is expected to have a typical intermediate-scale value. The effect of taking $\Lambda_{\rm susy}\sim\rho^{-1}$ is most noticeable in the case of $m_0=0$ which we consider here. On purely phenomenological grounds, the effect of considering $\Lambda_{\rm susy}$ as a new parameter in unified models
was (to our knowledge) first discussed in Ref.~\cite{KLN}.

Before dwelling into the low-energy phenomenology, let us elucidate the
value of $\Lambda_{\rm susy}\sim\rho^{-1}$ that we wish to consider. Following
Ref.~\cite{Witten} we set $M_{11}=M_{\rm LEP}\sim10^{16}\,{\rm GeV}$ in Eq.~(\ref{eq:M11def}). This allows us to solve for $V=M^{-6}_{\rm LEP}/2\alpha_G$. Substituting this value
in the expression for $\rho^{-1}$ in Eq.~(\ref{eq:rhodef}) we obtain
\begin{equation}
\rho^{-1}={4\over\alpha_G}\left({M_{\rm LEP}\over M_{\rm Pl}}\right)^3\,
M_{\rm Pl}\sim (10^{12}-10^{13})\,{\rm GeV}\ ,
\label{eq:rho}
\end{equation}
which is very suggestive in the context of gaugino condensation where
$m_{\rm susy}\sim \rho^{-3}/M^2_{\rm Pl}\sim M_Z$. It also follows that
\begin{equation}
R^{-1}=(2\alpha_G)^{1/d}\left({M_{\rm LEP}\over r^{-1}}\right)^{(6-d)/d}
M_{\rm LEP}
\label{eq:R}
\end{equation}
Note that in the reasonable case of $r^{-1}\sim M_{11}\sim M_{\rm LEP}$,
the $d$ dependence of $R^{-1}$ is greatly reduced and $R^{-1}\sim M_{\rm LEP}$
is also obtained.

\begin{figure}[p]
\vspace{6in}
\includegraphics{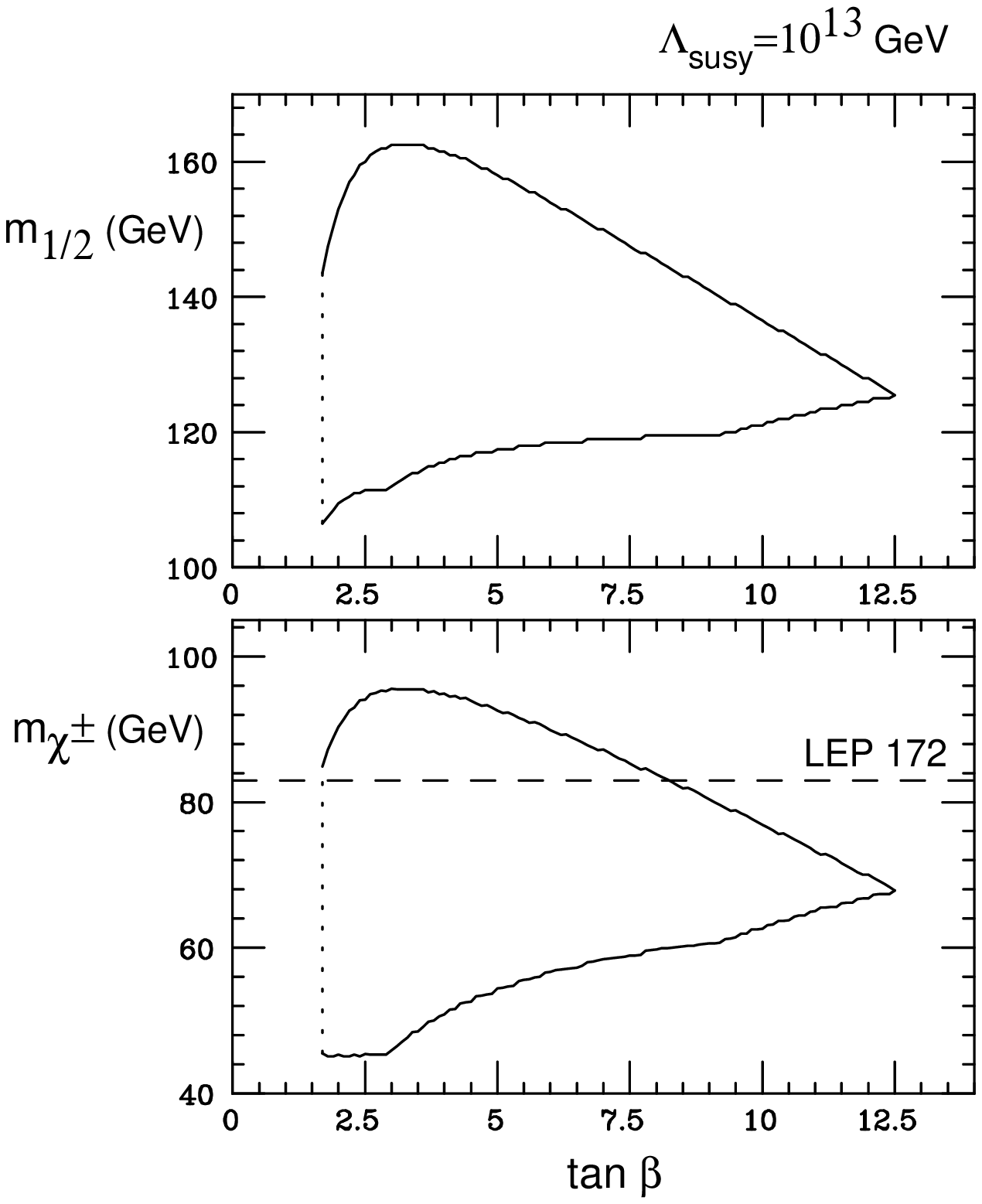}
\vspace{1cm}
\caption{The allowed region in $(m_{1/2},\tan\beta)$ [top panel] and
correspondingly $(m_{\chi^\pm},\tan\beta)$ [bottom panel] in no-scale
supergravity ($m_0=A_0=0$) with $\Lambda_{\rm susy}=10^{13}\,{\rm GeV}$.
Above the top boundary $m_{\tilde e_R}\approx m_{\tilde\tau_1}<m_\chi$, whereas below the bottom boundary $m_{\tilde\nu}<43\,{\rm GeV}$. The dashed line [bottom panel] represents the lower bound on the chargino mass from LEP~172 searches.}
\label{fig:boundary13}
\end{figure}

We now proceed to the analysis of the low-energy sparticle spectrum under the
assumptions of $\Lambda_{\rm susy}=\rho^{-1}$ and $m_0=A_0=0$.\footnote{The
procedure is standard, see {\em e.g.}, Refs.~\cite{KLN,Aspects}.} As remarked
above, the only free parameters are $m_{1/2}$ and $\tan\beta$. We find that
the requirement of radiative electroweak symmetry breaking plus two basic
phenomenological requirements, allow solutions in the $(m_{1/2},\tan\beta)$
plane for only one sign of $\mu$ and only within a completely bounded region.
For the case of $\Lambda_{\rm susy}=10^{13}\,{\rm GeV}$, this region is
shown in Fig.~\ref{fig:boundary13}, where to facilitate comparison with
experiment we also show the region in the $(m_{\chi^\pm},\tan\beta)$ plane. 
The upper limit on $m_{1/2}$ (for a fixed value of $\tan\beta$) follows from
the requirement that the lightest supersymmetric particle be neutral \cite{Aspects}. Above the upper boundary the right-handed selectron ($\tilde e_R$) becomes lighter than the lightest neutralino ($\chi$).\footnote{In practice, because we allow for a non-zero tau lepton mass, the lightest stau mass eigenstate ($\tilde\tau_1$) is in effect slightly lighter than the right-handed selectron.} The bottom boundary is obtained by imposing the absolute lower limit on the sneutrino mass from LEP~1 searches ($m_{\tilde\nu}>43\,{\rm GeV}$). The area to the right of the right-most tip of the region is excluded by these two conflicting constraints. The $\tan\beta$ dependence of these constraints may be understood from the D-term contribution to the $\tilde e_R$
and $\tilde\nu$ mass formulas
\begin{equation}
\widetilde m^2_i=c_i\, m^2_{1/2} - d_i\, {\tan^2\beta-1\over\tan^2\beta+1}\, M^2_W\ ,
\label{eq:masses}
\end{equation}
where the $c_i$ are some RGE-dependent constants and $d_{\tilde e_R}=-\tan^2\theta_W<0$ whereas $d_{\tilde\nu}={1\over2}(1+\tan^2\theta_W)>0$.
The dotted line indicates the lower bound on $\tan\beta$ that is consistent with the top-quark mass ($m_t=175\,{\rm GeV}$) and perturbative Yukawa couplings up to the unification scale \cite{Aspects}. In practice, the LEP~172 lower bound on the chargino mass ($m_{\chi^\pm}>83\,{\rm GeV}$ \cite{Nov19})\footnote{This limit applies as long as $m_{\chi^\pm}-(m_\chi\,{\rm or}\,m_{\tilde\nu})>3\,{\rm GeV}$, which is always satisfied in this model.} gives the strongest constraint on the parameter space (dashed line on bottom panel in Fig.~\ref{fig:boundary13}). Nonetheless, a portion of the parameter space remains allowed, and in fact it is within the reach of future LEP~2 energy upgrades, as we discuss below.

\begin{figure}[p]
\vspace{6in}
\includegraphics{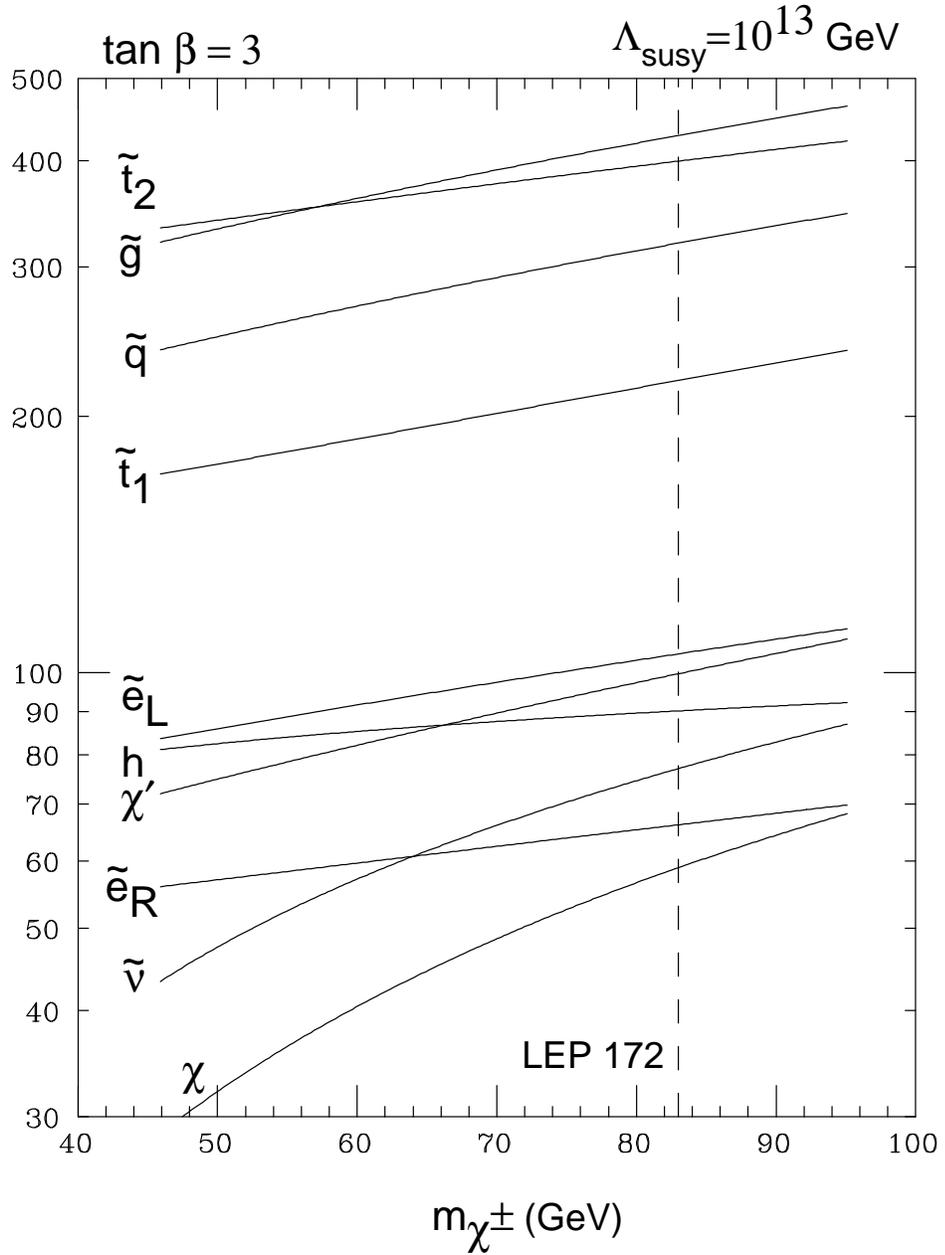}
\vspace{2cm}
\caption{Calculated values of representative sparticle masses versus the chargino mass for $\Lambda_{\rm susy}=10^{13}\,{\rm GeV}$ and $\tan\beta=3$.
The spectrum terminates when $m_\chi$ approaches $m_{\tilde e_R}$ from below.
The dashed line represents the lower bound on the chargino mass from LEP~172 searches.}
\label{fig:spectrum}
\end{figure}

To give a more detailed picture of the low-energy spectrum, in Fig.~\ref{fig:spectrum} we display representative sparticle masses as a function of the chargino mass for $\Lambda_{\rm susy}=10^{13}\,{\rm GeV}$ and
$\tan\beta=3$. This choice of $\tan\beta$ allows the widest range of sparticle
masses (see Fig.~\ref{fig:boundary13}). This figure shows that the spectrum
`terminates' when $m_\chi$ approaches $m_{\tilde e_R}$ from below, as mentioned
above in connection with the upper boundary in Fig.~\ref{fig:boundary13}. It is
interesting to note the significant splitting of the top-squark ($\tilde t_{1,2}$) masses around the average squark ($\tilde q$) mass.

In the LEP 172 allowed region in Figs.~\ref{fig:boundary13} and \ref{fig:spectrum} we find $m_{\chi^\pm_1}<95\,{\rm GeV}$ and $m_{\tilde e_R}<70\,{\rm GeV}$. Both of these particles appear within the reach of LEP~2. More to the point, one might
wonder whether the such light right-handed selectron masses might have already
been excluded by LEP~2 searches, as they have been certainly kinematically accessible. We have calculated the cross section $\sigma(e^+e^-\to\tilde e^+_R
\tilde e^-_R)$ at LEP~161, for which explicit limits have been released by
the OPAL Collaboration \cite{OPAL}. We find $\sigma<0.2\,{\rm pb}$, which in
${\cal L}=10.1\,{\rm pb}^{-1}$ would have yielded a maximum of two events.
Indeed, the experimental sensitivity to this mode is at the 0.5~pb level \cite{OPAL}. Thus, past LEP~2 searches in the selectron channels do not restrict the allowed parameter space any further.\footnote{Moreover, near the
upper end of the parameter space the experimental detection efficiency should
be greatly reduced because $m_{\tilde e_R}$ approaches $m_\chi$.} One should also consider the predictions for trilepton events at the Tevatron. We find $\sigma(p\bar p\to\chi^\pm\chi')\approx (1.0-0.7)\,{\rm pb}$ for $m_{\chi^\pm}=(83-95)\,{\rm GeV}$. The leptonic decays of the chargino and
neutralino are maximally enhanced because of the lighter right-handed sleptons
and sneutrinos, respectively. That is, $B(\chi^\pm\to\ell\nu_\ell\chi)\approx2/3$
and $B(\chi'\to\ell^+\ell^-\chi)\approx1/2$, where $\ell=e+\mu$. Combining these numbers we arrive at a single channel ({\em i.e.}, any single one of
$eee$, $ee\mu$, $e\mu\mu$, or $\mu\mu\mu$) cross section of $(0.16-0.11)$~pb.
This result is slightly below the sensitivity reached at the Tevatron in
trilepton searches \cite{Wyatt}, and thus these also do not constrain the
allowed parameter space any further.

\begin{figure}[p]
\vspace{6in}
\includegraphics{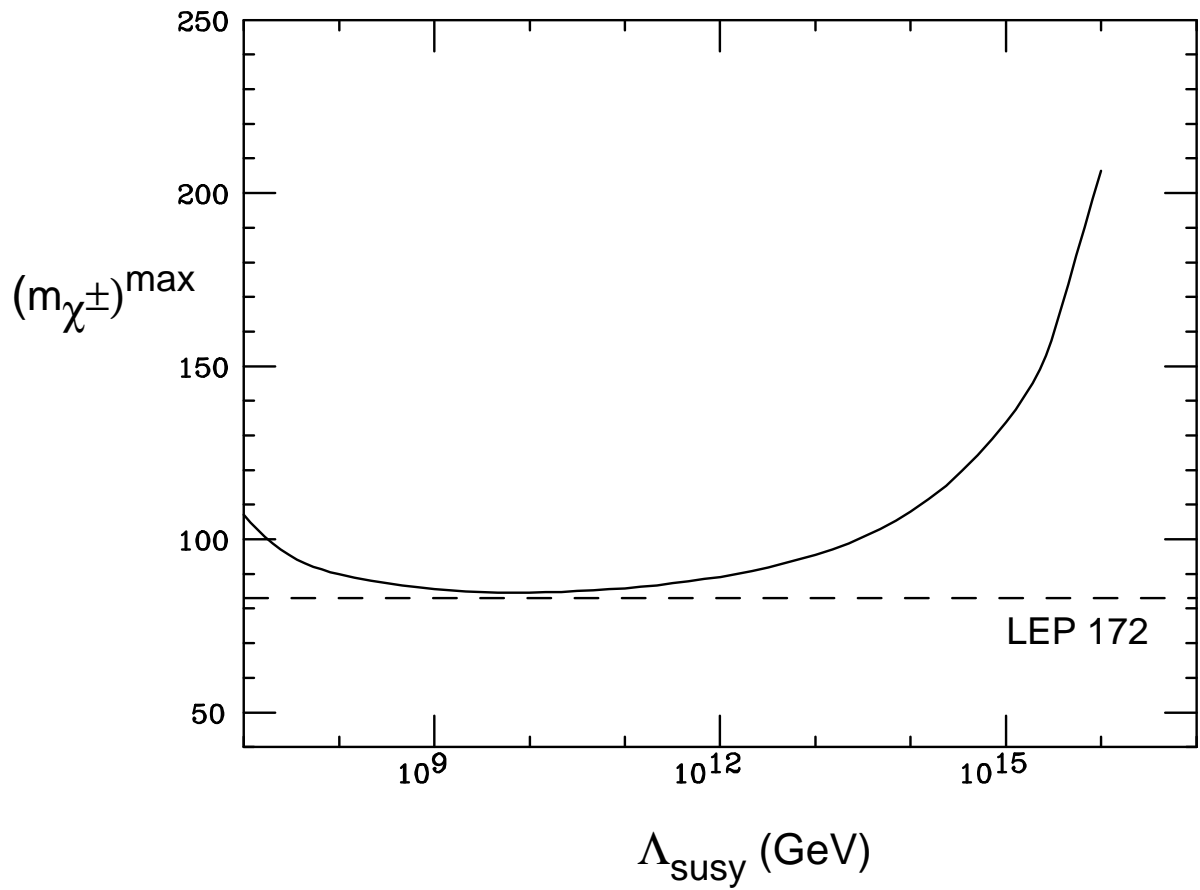}
\caption{The maximum allowed value of the chargino mass as a function of
$\Lambda_{\rm susy}$ in no-scale supergravity ($m_0=A_0=0$). The dashed line represents the lower bound on the chargino mass from LEP~172 searches.}
\label{fig:chmax}
\end{figure}

One may also want to study the effect of different choices for $\Lambda_{\rm susy}$. These affect the shape of the bounded region in Fig.~\ref{fig:boundary13}, although not very significantly in the interval
specified by Eq.~(\ref{eq:rho}). As a means of quantifying this behavior, one can determine the maximum allowed ({\em i.e.}, passing LEP~1 cuts) chargino masses for a range of $\Lambda_{\rm susy}$ values. These are shown in Fig.~\ref{fig:chmax}. The figure shows that LEP~2 data presently do not constrain $\Lambda_{\rm susy}$ in any way, although future higher energy runs will soon begin to constrain $\Lambda_{\rm susy}$.

\section*{Acknowledgments}
This work has been partially supported by the World Laboratory. The work of J.~L. has been supported in part by DOE grant DE-FG05-93-ER-40717 and that of D.V.N. by DOE grant DE-FG05-91-ER-40633.

\clearpage

\clearpage

\clearpage

\end{document}